\documentclass[article]{aa}
\usepackage{graphicx}
\usepackage{txfonts}
\usepackage{xcolor}
\usepackage[hidelinks]{hyperref}
\hypersetup{
  colorlinks,
  linkcolor={red},
  citecolor={blue!80!black},
  urlcolor={blue!80!black}
}
\usepackage{color}
\usepackage{siunitx}

\begin{document}

\title{Beyond the exoplanet mass-radius relation \thanks{Data sets are available in electronic form at the CDS via anonymous ftp to \href{cdsarc.u-strasbg.fr}{cdsarc.u-strasbg.fr (130.79.128.5)} or via \href{http://cdsweb.u-strasbg.fr/cgi-bin/qcat?J/A+A/}{http://cdsweb.u-strasbg.fr/cgi-bin/qcat?J/A+A/}}}
\titlerunning{ }
\author{S. Ulmer-Moll\inst{\ref{inst1},\ref{inst2}},
  N.\,C. Santos\inst{\ref{inst1},\ref{inst2}},
  P. Figueira\inst{\ref{inst3}, \ref{inst1}},
  J. Brinchmann\inst{\ref{inst4}, \ref{inst1}},
  and J.\,P. Faria\inst{\ref{inst1},\ref{inst2}}
}

\institute{
  Instituto\,de\,Astrofísica\,e\,Ciências\,do\,Espaço,\,Universidade\,do\,Porto,\,CAUP,\,Rua\,das\,Estrelas,\,4150-762\,Porto,\,Portugal\label{inst1}
  \and
  Departamento de Física e Astronomia, Faculdade de Ciências, Universidade do Porto, Portugal\label{inst2}
  \and
  European Southern Observatory, Alonso de Cordova 3107, Vitacura, Santiago, Chile\label{inst3}
  \and
  Leiden Observatory, Leiden University, Leiden, The Netherlands\label{inst4}}

\date{Received 11 June 2019 / Accepted 30 August 2019}

\abstract
    {Mass and radius are two fundamental properties for characterising exoplanets, but only for a relatively small fraction of exoplanets are they both available.
      Mass is often derived from radial velocity measurements, while the radius is almost always measured using
      the transit method.
      For a large number of exoplanets, either the radius or the mass is unknown, while the host star has been characterised.
      Several mass-radius relations that are dependent on the planet's type have been published that often allow us to predict the radius.
      The same is true for a bayesian code, which forecasts the radius of an exoplanet given the mass or vice versa.}
    {Our goal is to derive the radius of exoplanets using only observables extracted from spectra
          used primarily to determine radial velocities and spectral parameters.
      Our objective is to obtain a mass-radius relation independent of the planet's type.}
    {We worked with a database of confirmed exoplanets with known radii and masses, as well as the planets from our Solar System.
      Using random forests, a machine learning algorithm, we computed the radius of exoplanets and compared the results to the published radii.
      Our code, \href{https://github.com/soleneulmer/bem}{\texttt{Bem}}, is available online.
      In addition, we explored how the radius estimates compare to previously published mass-radius relations.}
    { The estimated radii reproduces the spread in radius found for high mass planets better than previous mass-radius relations.
      The average radius error is 1.8$\,R_\oplus$ across the whole range of radii from 1-$22\,R_\oplus$.
      We find that a random forest algorithm is able to derive reliable radii,
      especially for planets between $4\,R_\oplus$ and $20\,R_\oplus$
      for which the error is under  $25\%$.
      The algorithm has a low bias yet a high variance, which could be reduced by limiting the growth of the forest, or adding more data.}
    {The random forest algorithm is a promising method for deriving exoplanet properties.
      We show that the exoplanet's mass and equilibrium temperature are the relevant properties that constrain the radius,
      and do so with higher accuracy than the previous methods.}


  \maketitle

%

  \section{Introduction}

  Mass and radius are two fundamental parameters for characterising exoplanets.
  The two most prolific methods to detect exoplanets are the transit method (e.g. \citealt{deegTransitPhotometryExoplanet2018a})
  and the radial velocity method (e.g. \citealt{wrightRadialVelocitiesExoplanet2017}),
  which give access to different parameters.
  The mass is derived through the radial velocity method, while the radius is measured using the transit method.
  These two parameters may be obtained via other methods.
  The mass can be derived via microlensing,
  and the radius, while degenerate with other parameters, from direct detection.
  Time transit measurements allow one to determine planetary masses through gravitation interaction \citep{becker_2015},
  but these remain a minority.
 
  Several previous works demonstrated that the relation between the mass and radius of a gravitationally bound object
  can be described with a polytropic relation \citep{burrows_1993,chabrier_2000,chabrier_2009}.
  Mass-radius relations depending on the planetary composition
  have been produced in order to infer the structure of exoplanets \citep{seager_2007,swift_2011}.

  \cite{weiss_mass_2013}, propose that the mass-radius relation of planets can be explained by two power laws,
  one for low-mass planets ($< 150\,M_{\oplus}$) and another for high-mass planets ($> 150\,M_{\oplus}$).
  Following this work, \cite{bashi_two_2017} propose a revised version of the power law exponents
  with a breakpoint between the two mass regimes at $124 \pm 7\,M_\oplus$.
  \cite{hatzes_definition_2015a} present a mass-density relation
  divided into three areas supported by underlying physics: low mass planets ($< 95\,M_{\oplus}$),
  gas giant planets ($< 60\,M_{J}$), and stellar objects ($> 60\,M_{J}$).
  While these parametric relations draw the general trend of the mass-radius relation,
  they are limited in their ability to explain
  the spread of exoplanet radii at fixed mass.
  The fixed mass limits are sometimes defined in an ad hoc way.

  \citet{wolfgang_2016a} introduce a probabilistic model for the mass-radius relation for small exoplanets (< 8\,$R_\oplus$)
  assuming a power law description of the relation.
  \citet{chen_2017a} extend this idea to a larger data set,
  predicting the mass or the radius of an astronomical object over four orders of magnitude.
  In a follow-up paper, the authors computed the predicted mass for over 7000 Kepler Objects of Interest \citep{chen_forecasted_2018}.
  Their code, \texttt{Forecaster},
  is available to the community\footnote{\href{https://github.com/chenjj2/forecaster}{github.com/chenjj2/forecaster}}
  and is able to reproduce a larger spread in radius (or mass) than the previous power law relations.

  For all the methods presented, the transitional points are either fixed or fitted
  in order to describe the variety of objects covering a range of masses of one or more order of magnitude.
  However, while there is no inherent problem in trying to classify the astronomical objects,
  there is no consensus on the number of classes (i.e. number of power laws) chosen to describe the mass-radius relation.
  To avoid these caveats, \citet{ning_2018} present a non-parametric
  approach\footnote{\href{https://github.com/Bo-Ning/Predicting-exoplanet-mass-and-radius-relationship}{github.com/Bo-Ning/Predicting-mass-radius}}
  to model the mass-radius relation for small exoplanets, with the sample of \citet{wolfgang_2016a}.
  \cite{kanodia_2019} used the same non-parametric method to predict the radius of 24 exoplanets
  orbiting M dwarf stars.
  Their code \texttt{MRexo} is also available online\footnote{\href{https://github.com/shbhuk/mrexo}{github.com/shbhuk/mrexo}}.
  
  The radius of giant planets has also been correlated to other physical parameters
  such as the equilibrium temperature, the orbital semi major axis, the tidal heating rate,
  the stellar metallicity, and stellar irradiation \citep{guillot_2006,fortney_2007,enoch_factors_2012}.
  For different classes of giant exoplanets,
  \cite{bhatti_hats19b_2016} used a random forest model
  to demonstrate the influence of equilibrium temperature and planetary mass
  on the radius estimation.

  We propose using an algorithm that does not require a previous classification of the objects to do the predictive work.
  The growing number of exoplanets with mass and radius measurements
  allows one to look at machine learning techniques to model the mass-radius relation.
  We focus on the prediction of exoplanet radii using observables extracted from spectra and stellar parameters
  by using a random forest model akin to that of \cite{bhatti_hats19b_2016}.
  In Section \ref{sample}, we introduce the sample and the description of the modelling tool we used.
  Section \ref{results} features the results obtained with the random forest model,
  \href{https://github.com/soleneulmer/bem}{\texttt{Bem}}, and a comparison with previous works.
  In Section \ref{conclusion}, we summarise our findings, and
  we discuss the improvements that will benefit the predictive work of the algorithm.

%

\section{Data and methods}
\label{sample}

We present the data set of exoplanets in section~\ref{sample1}.
The random forest model is explained in section~\ref{rf}, and the
selection of the features, used as input for the model, is presented in section~\ref{feature}.

\subsection{Sample selection}
\label{sample1}
We selected exoplanets with mass and radius measurements,
as well as planetary and stellar parameters, which can be derived from radial velocity
and spectral analyses respectively. We collected a total of 501 exoplanets
to which we added the eight planets of the Solar System.
We obtained the parameters of the discovered exoplanets and the host stars from
\textit{The Extrasolar Planets Encyclopaedia}\footnote{\href{http://exoplanet.eu/catalog/}{exoplanet.eu/catalog}} on April 15, 2019, the parameters for the solar system planets, and one Kuiper Belt object
from \textit{Planetary Fact Sheet}\footnote{\href{https://nssdc.gsfc.nasa.gov/planetary/factsheet/index.html}{nssdc.gsfc.nasa.gov/planetary/factsheet/index.html}}.
We removed three exoplanets because of their unreliable mass measurements (HATS-12 b, K2-95 b, and Kepler-11 g)
\footnote{HATS-12\,b: its planetary mass may decrease due to the large decrease in the host star luminosity \citep{johns2018}.
    K2-95\,b: the radial velocity observations only placed an upper limit on the planetary mass \citep{pepper_2017}.
    Kepler-11\,g: the planetary mass is only constrained by upper bounds \citep{lissauer_2013,bedell_2017}.}.
We computed the equilibrium temperature of the exoplanets
following Equation 5 from \cite{laughlinExoplanetaryGeophysicsEmerging2015},
without taking into account the effect of the albedo.
We also included the redistribution factor presented
in Equation 3.9 from \cite{seagerExoplanetAtmospheresPhysical2010}.
As the author explains, the redistribution factor is equal to 1/4,
assuming that the stellar radiation is uniformly distributed around the exoplanet.

\begin{figure}
  \includegraphics[width=\hsize]{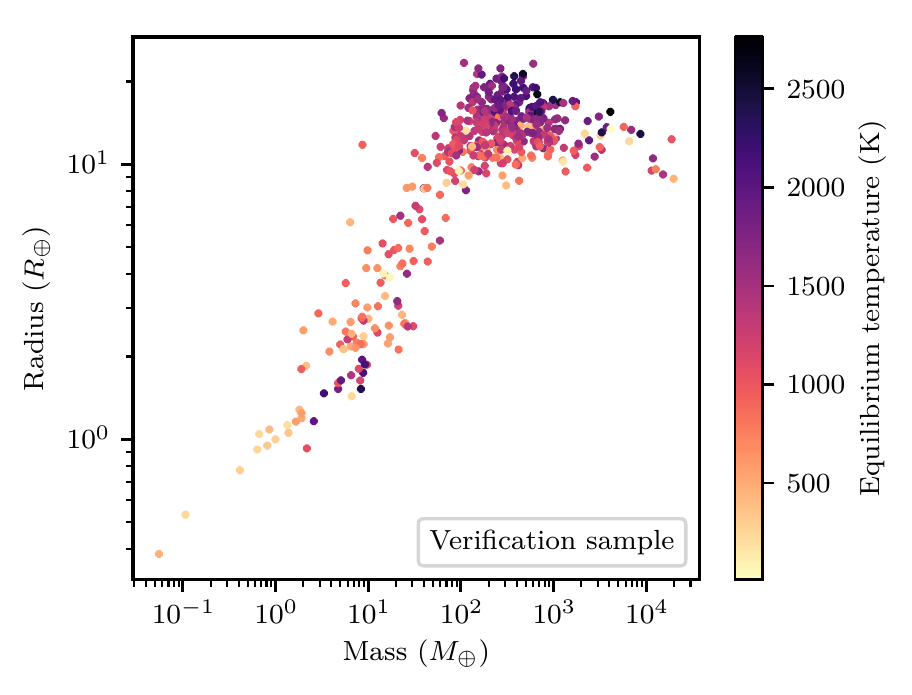}
  \caption{Sample of selected exoplanets with radius measurements plotted as a function of mass and equilibrium temperature.}
  \label{fig:sample}
\end{figure}

We present the sample of exoplanets in Figure~\ref{fig:sample}.
In a lower mass regime, this figure shows the clear positive correlation between mass and radius.
The planets with lower mass have a smaller radius.
For planets with masses larger than $10^2\,M_{\oplus}$,
higher equilibrium temperature is associated with larger radius.
We also notice a group of exoplanets (masses from 3-11\,$M_{\oplus}$)
with high equilibrium temperature and small radius.
A hypothesis is that these close-in planets, with a semi-major-axis smaller than 0.02 AU,
could have undergone evaporation due to their proximity to the host star \citep{lammer_2003}.
In this plot, we do not define one or several transition masses to separate
the low-mass and high-mass planets.
We pinpoint general trends, which are not explicitly included in the machine learning algorithm,
but they will help define the parameters used as input features, such as planetary mass.

\subsection{Random forest algorithm}
\label{rf}
Random forest, introduced by \cite{breiman_random_2001}, is a predictive modelling tool.
This methodology consists of extracting information from existing data sets, 
and possibly uncovering new correlations in order to predict a variable.
In our case, we used random forest to perform a regression task
and predict the radius of an exoplanet when given other observables.
We can look at the importance of the different planetary and stellar parameters
and explore how each parameter impacts the predicted radii.
However, the random forest modelling does not allow us to write down
a parametric relation between the radius and the other parameters.

Random forest is an ensemble method. In order to provide the final radius estimate,
random forest combines the results of several estimators called decision trees.
A decision tree is an algorithm that classifies an object as a label
by checking several conditions.
A decision tree is composed of nodes (cells where a statement has to be checked)
and leaves (cells with a radius estimate).
A branch begins at the first node of the tree and continues to one of its leaves.

An ensemble of decision trees composes a random forest.
The decision trees have a variable number of leaves.
Parameters including the number of trees and the number of leaves in a random forest are called hyperparameters.
Hyperparameters can be changed by the user or automatically explored with algorithms such as random search and grid search. In our case, we used a grid search to optimise the value of the hyperparameters.
The random forest algorithm was first built on a training set, which usually contains 70-90\% of the total data set
\citep{guyon_1997,louppeUnderstandingRandomForests2014}.
The rest of the data set, known as a test set, was reserved to compare its result with the result obtained with the training set.
Once the value of the hyperparameters were found,
we applied the random forest with fixed hyperparameters to the test set.

\subsection{Feature selection}
\label{feature}

Feature selection is the selection of relevant observables to be used
with a machine learning algorithm and it is
an essential preliminary step to perform before using the random forest regression.
We started with the fundamental parameters of the exoplanet and exoplanet's orbit,
then we added the stellar parameters, and finally we added parameters computed from the planetary and stellar parameters.
All the features are taken from or computed with parameters from \textit{The Extrasolar Planets Encyclopaedia.}
Initially, we chose ten features to train the random forest algorithm.
The planet's parameters are
mass, equilibrium temperature, semi major axis, orbital period, and eccentricity.
The star's parameters are radius, effective temperature, mass, metallicity, and luminosity.
All these parameters are physically motivated,
as they are all thought to be able to or
have been shown to play a role in the mass-radius relation \citep{enoch_factors_2012}.
We ran the random forest algorithm with several ranges of hyperparameters
and we checked the feature importance to refine the selected features.
We computed the importance of the features using the mean decrease impurity
\citep{breiman_random_2001,louppeUnderstandingRandomForests2014}
as implemented in Scikit-learn \citep{pedregosa_scikitlearn_2011}.
The impurity was evaluated in each node,
and it can be seen as a measure of the similarity of the exoplanets in the node.
The impurity is at its lowest when all the exoplanets in the node have similar radii.
The mean decrease impurity is the ratio
between the decrease in  node impurity and the probability of reaching that node.
The radius predictions did not improve when adding the three least important features, so
we kept the seven most important features:
the planet's mass, equilibrium temperature, semi major axis,
stellar luminosity, mass, radius, and effective temperature.

Random forest predictors can be subject to high variance, which means that,
using the same features, different training sets will result in different models.
To reduce the variance,
we choose to use the random subspace method, also known as feature bagging \citep{tinkamho_1998}.
The feature bagging technique allows one to reduce the correlation between decision trees.
In our case, the planetary mass was the best predictor of the planetary radius.
The mass feature tends to be chosen more often than other features
to perform the splitting of the data, resulting in correlated trees \citep{hastie_elements_2009}.
To perform feature bagging, we limited the feature space
from which a feature can be selected to split the data in a node.
At each split in a decision tree, the feature selected is taken from a random subspace of four features,
instead of the full feature space composed of seven features.
We also chose to set the minimum node size of four,
which means each node needs to contain more than four samples to be split.
Both methods, the feature bagging and the minimum node size,
were designed to reduce the variance of random forests.

\section{Results}
\label{results}

We used two samples to test the random forest algorithm.
The first sample contains mass and radius measurements for all its objects:
the exoplanets and the Solar System objects.
This verification sample is randomly split into a training set and a test set.
The results of predicted radii on the test set are presented in Section \ref{verif}.
We explain in detail the radius predictions for five exoplanets in Section \ref{lime},
and discuss these results in Section \ref{discussion}.
The second sample is composed of exoplanets discovered by the radial velocity method
and without a radius measurement. The results are presented in Section \ref{rv_sample}.

\subsection{Verification sample}
\label{verif}

The verification sample is composed of 506 objects with mass and radius measurements,
as described in section \ref{sample1} and shown in Figure~\ref{fig:sample}.
We designed a training set composed of 75\% of the verification sample,
and the remaining 25\% of objects forms the test set.
Both sets are available at the Centre de Donn\'{e}es astronomiques de Strasbourg (CDS) as two tables containing the following information.
Column 1 lists the names of the planets, columns 2 and 3 contain the planetary mass and radius,
column 4 gives the semi-major-axis, and column 5 presents the planetary equilibrium temperature.
Columns 6, 7, 8, and 9 give the stellar luminosity, mass, radius, and effective temperature respectively.
The variable hyperparameters of our random forest model are the number of trees, the depth of the trees,
and the number of features available to split a node (feature bagging).
To build the random forest, we used the estimator \texttt{'RandomForestRegressor'} and
the cross validated grid search method \texttt{'GridSearchCV'} from Scikit-learn \citep{pedregosa_scikitlearn_2011}.

\begin{figure}
  \includegraphics[width=\hsize]{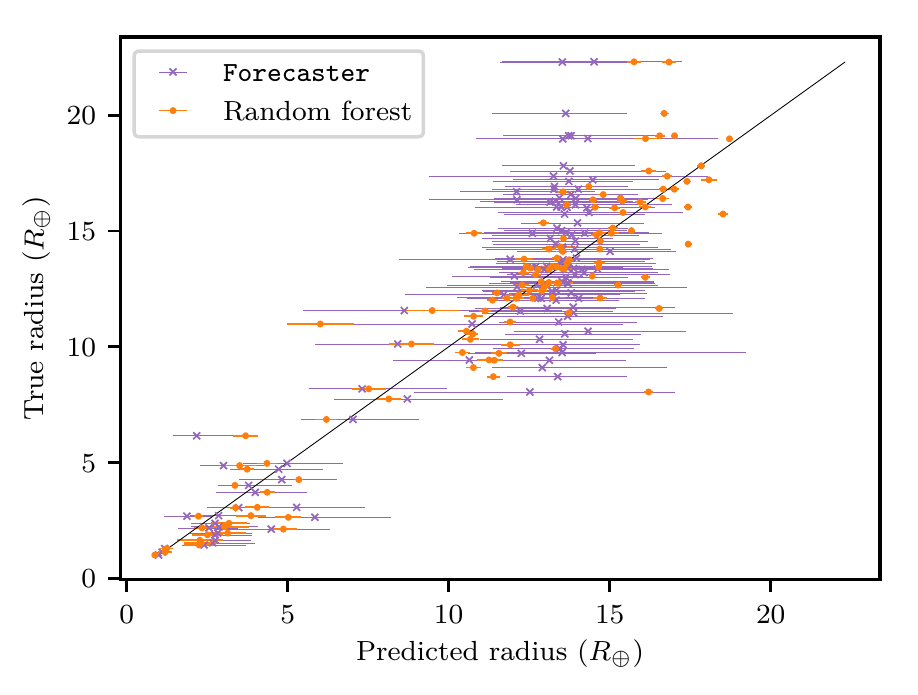}
  \caption{True radii as a function of predicted radii for test set.
    Radii obtained with the random forest algorithm (orange dots) and
    \texttt{the Forecaster} code (purple crosses) are compared with the 1:1 line in black.}
  \label{fig:test_radii}
\end{figure}

The random forest model is built with the training set, and
the radii of both the training and test sets are predicted with the model.
The root-mean-squared error on the radius prediction is 1.1$\, R_\oplus$ for the training set,
and 1.8$\,R_\oplus$ for the test set.
For comparison, the average radius uncertainty found in the sample is $0.6\, R_\oplus$,
so we consider the error on the training set to be small.
However, the lower error on the training set, relative to the test set,
shows that the training set tends to be overfitted,
even though the difference between the two errors remains small.
To evaluate the quality of the radius predictions ($\hat{y}$)
compared to the radius measurements ($y$),
we used the coefficient of determination implemented in Scikit-learn, also known as the $R^2$ score.
The $R^2$ score is defined as follows:
\[ R^2(y, \hat{y}) = 1 - \frac{\sum_{i=0}^{n_{\text{samples}} - 1} (y_i - \hat{y}_i)^2}{\sum_{i=0}^{n_\text{samples} - 1} (y_i - \bar{y})^2}
$  with  $ \bar{y} =  \frac{1}{n_{\text{samples}}} \sum_{i=0}^{n_{\text{samples}} - 1} y_i. \]
The $R^2$ score can be negative when the  prediction error is larger than the error relative to the mean
and the best $R^2$ score is one, which corresponds to perfect prediction.
For the test set, the $R^2$ score is equal to 0.87,
and the Pearson correlation coefficient between $y$  and $\bar{y}$ is equal to 0.93.
We calculated the importance of the feature with the function implemented in Scikit-learn.
The importance of the features demonstrates that the planet's mass is clearly
the most important parameter, followed by the planet's equilibrium temperature.
The stellar parameters and the semi major axis are
the features that are the least important in predicting the planetary radius.

\begin{figure}
  \includegraphics[width=\hsize]{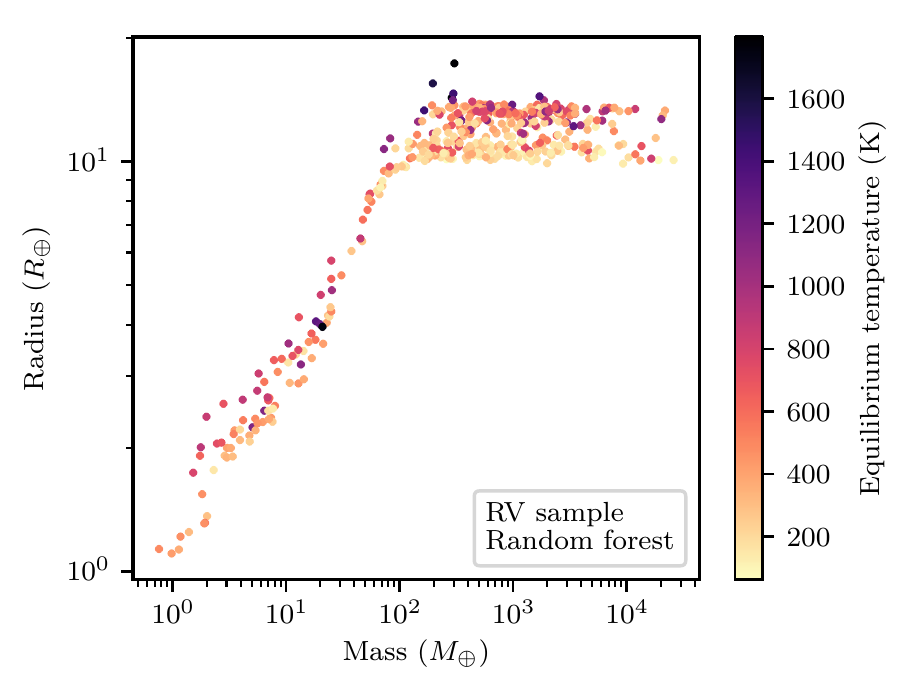}
  \caption{Predicted radii as a function of mass for radial velocity sample.
    Radii obtained with the random forest algorithm.}
  \label{fig:rv_RF}
\end{figure}

For the 127 exoplanets in the test set, Figure~\ref{fig:test_radii} presents the predicted radii
by the random forest algorithm together with the \texttt{Forecaster} radii
as compared to the radius measurements from the database.
The error bars for the random forest model are computed using a Monte Carlo approach.
For each feature, an updated value is drawn from a normal distribution
centred on the original value with a standard deviation equal to the uncertainty.
If the uncertainty is not defined,
the standard deviation is set to the 0.9  quantile
of the distribution of uncertainties for each feature.
The radius is predicted using the same model without training it again.
The root-mean-squared error is around 1.8$\,R_\oplus$ for the random forest model,
and 2.5$\,R_\oplus$ for \texttt{Forecaster}.
The \texttt{Forecaster} predictions for planets
with radius between $10\,R_\oplus$ and $20\,R_\oplus$ tend to cluster around a predicted radius of 13-$14\,R_\oplus$.
The figure shows that both models have a low bias but a large variance compared to the 1:1 line.

For the 26 exoplanets in the test set that have a radius under 8\,$R_\oplus$,
we compared the radius prediction of the random forest model with the non-parametric model \texttt{MRExo}
\citep{ning_2018,kanodia_2019}. We predicted that the exoplanets with a host star mass smaller than 0.6\,$M_\odot$
to be part of the M dwarf sample and the other planets to be part of the Kepler sample. 
The root-mean-squared error is equal to 1.3\,$R_\oplus$ for \texttt{MRExo}, and to 1.1\,$R_\oplus$ for the random forest model.

\begin{figure}
  \includegraphics[width=\hsize]{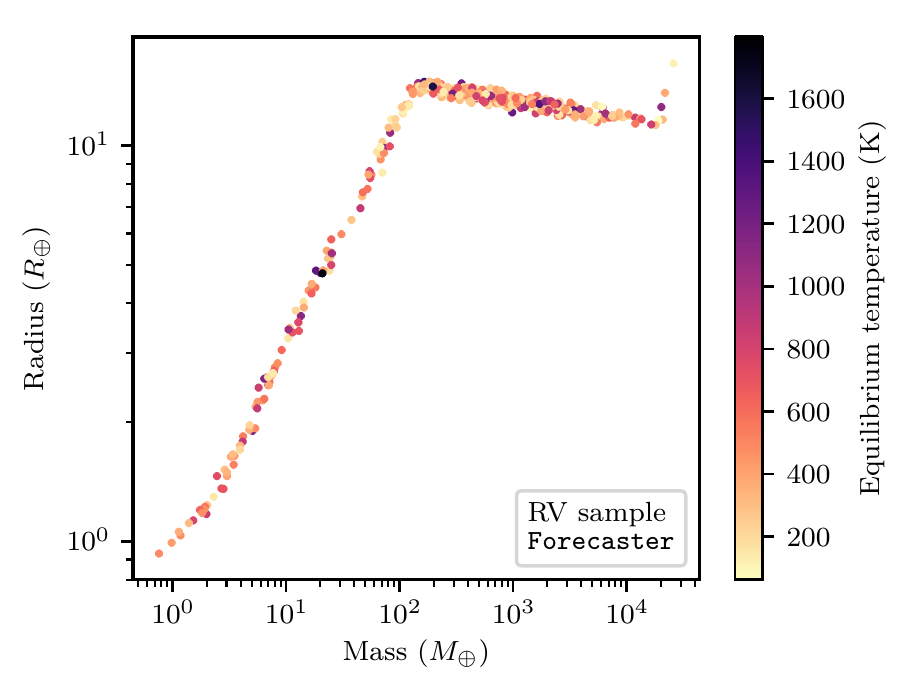}
  \caption{Predicted radii as a function of mass for radial velocity sample.
    Radii obtained with the \texttt{Forecaster} code.}
  \label{fig:rv_Forecaster}
\end{figure}

The learning and validation curves presented in Appendix~\ref{fig:curves},
allow the diagnosis of the random forest model.
In general terms, we find that this model has a low bias and a high variance.
The high variance means that when another training set
is used to build the random forest model, the estimated radii changes.
In other words, the radius predictions are accurate but have a low precision.
Since we already implemented methods like feature bagging to reduce the variance of the model,
adding more training samples will improve our model.

\subsection{Interpretation of radius predictions}
\label{lime}

To interpret the predictions of the random forest model, we used the
Local Interpretable Model-agnostic Explanations (LIME) technique \citep{ribeiro_why_2016}.
In our context, LIME approximates locally, with a linear function,
a particular exoplanet for which we want to explain the radius prediction.
LIME returns the input features and their relative weights
used to predict the radius of the exoplanet
and allows one to locally visualise  how the features influence the radius prediction.

In Appendix~\ref{fig:lime}, we present five exoplanets from the test set:
the radius is well-predicted for three of them, and the other two are wrongly predicted by the random forest model.
These five planets were chosen as examples to demonstrate particular cases of (mis-)prediction.

HATS-35\,b is a moderately inflated hot Jupiter with a radius of 16.4$R_\oplus$ \citep{deval-borro2016}.
Our model predicts a radius of 16.5$R_\oplus$.
The LIME approximation shows that all the input features have a positive weight,
which tends to increase the radius.
This is an expected outcome, because the exoplanet is massive and has a high equilibrium temperature.

However, WASP-17\,b is a highly inflated hot Jupiter with a radius of 22.3$R_\oplus$ \citep{anderson_2011}.
In this case, the model underestimates the radius, with a predicted radius of 15.7$R_\oplus$.
All the features have a positive weight, which is a reasonable behaviour for the model.
But WASP-17\,b is one of the largest exoplanets discovered to date,
and there are few exoplanets of this nature in the training sample. 

CoRoT-13\,b is a dense exoplanet with a large amount of heavy elements \citep{cabrera_2010}.
Its radius of 9.9$R_\oplus$ is overestimated by our model, which predicts 13.3$R_\oplus$.
While the equilibrium temperature, the stellar luminosity,
and the stellar radius push towards a smaller radius,
the positive effect of the mass is not compensated.
The presence of heavy elements can lead to a smaller radius
compared to the radius of exoplanets with atmospheres dominated by hydrogen and helium
\citep{guillot_2006,seager_2011}.
The fact that stellar metallicity is not considered in our model,
because it had a relatively small role in defining the radius for most of the planets,
may be the cause for the observed offset.

Kepler-75\,b has a radius of 11.5$R_\oplus,$ \citep{hebrard_2013}
which is well estimated by the model as 11.0$R_\oplus$.
The high mass of the exoplanet gives a positive weight,
which is compensated by the negative weights of all the other features.
The exoplanet's low equilibrium temperature has a large negative weight,
which could be the reason why the radius is well predicted even though the exoplanet is massive.

Finally, Kepler-20\,c has a radius of 3.1$R_\oplus$ \citep{gautier_2012}.
The predicted radius of 3.4$R_\oplus$ is close to the measured radius.
Almost all the features have negative weights, which is the expected behaviour for the model.

\subsection{Effect on radius predictions}
\label{discussion}

With LIME, we are able to analyse  the behaviour of the random forest model further.
We check the factors which affect the predicted radii.
We find that larger planetary mass and a higher equilibrium temperature
increase the planetary radius (and respectively for smaller radii).
These correlations were expected from previous mass-radius relations (e.g. \citealt{enoch_factors_2012})
and have been uncovered by the random forest model.

We also find that stellar parameters with higher values (e.g. large stellar mass)
lead to a larger planetary radius (and respectively for smaller radii).
These trends with the stellar parameters are, in part,
the result of naturally correlated quantities, such as effective temperature and luminosity.
But they are also the result of observational biases,
since we do not correct the input data set for any detection biases.
As explained in \cite{bhatti_hats19b_2016},
the radius of transiting giant exoplanets is correlated with the stellar mass.
The larger planets are easier to detect around bright stars that have larger luminosity, mass, and radius. 
On the other hand, \cite{jiangPlanetaryMassStellar2018} reports a positive trend
between the stellar radius and the planetary mass for a sample of red giant stars.

Another selection effect of the detection biases affects the exoplanet sample.
The exoplanets with mass and radius measurements are usually those
that satisfy the detection limits of both the transit and radial velocity methods.
The exoplanets that are easier to detect
with radial velocities, such as short-period planets,
are over-represented in the training sample.
This results in an imbalanced training sample.
\citet{chen_2004} present two ways to correct this imbalance,
one called 'sampling technique' (under or over-sampling) and the other 'cost-sensitive learning'.
For example, in our case, the first technique would imply selecting a smaller number of exoplanets
(under-sampling) that are over-represented, but this can result in a loss of information for this class of planets.
The second technique would attribute a larger weight for planets that are under-represented.
The final error increases more when the under-represented group
is wrongly predicted rather than the over-represented group.

Comparing our results to \cite{bhatti_hats19b_2016},
we also find that the radii of hot-Saturns ($32<M_p<159M_\oplus$) is primarily dependent on the planetary mass
followed by the equilibrium temperature. For the hot-Jupiters ($159<M_p<636M_\oplus$) and higher mass planets ($>636M_\oplus$),
the authors find that the radius depends mainly on the equilibrium temperature.
But, we find that for both groups, the planetary mass is still the main driver of the radius prediction
followed by the equilibrium temperature.
For the higher mass planets, the equilibrium temperature is the feature with the highest weight
in 40\% of cases in the test set.

It should be noted that to calculate the equilibrium temperature, we set the albedo to zero,
since few exoplanets have a measurement of their albedo.
This is a common approximation for hot Jupiters \citep{madhusudhan_exoplanetary_2014}.
For terrestrial planets, an albedo of zero or around 0.3, close to the Earth's value, is usually chosen.
However, this nominal value does not represent the variety of albedos
for terrestrial or potentially habitable exoplanets \citep{delgenio_2018}.
This approximation on the albedo probably has an impact on the radius prediction.

\subsection{Radial velocity sample}
\label{rv_sample}

The radial velocity sample is composed of 488 exoplanets collected from \textit{The Extrasolar Planets Encyclopaedia},
which have been discovered with the radial velocity method and do not have a radius measurement in the database.
Given the measured masses and stellar parameters, we
can make predictions about their radii and compare them with the \texttt{Forecaster} prediction.
We used the same random forest model as built with the verification sample.
Figure~\ref{fig:rv_RF} presents the predicted radii as a function of the mass and equilibrium temperature.
For high mass planets ($>10^{2}M_\oplus$), the gradient in equilibrium temperature
is well- estimated and results in a spread in radius for the same mass. 
For lower-mass planets, the mass-radius relation is tighter,
and the predicted radii appear to concentrate between $2\,R_\oplus$ and $4\,R_\oplus$.
We compare these results with the predicted radii with the \texttt{Forecaster}
model, which is shown in Figure~\ref{fig:rv_Forecaster}.
The predicted radii from \texttt{Forecaster} do not recover
the observed gradient with equilibrium temperature.
The mass-radius relation for all planets has a smaller spread in radius
than with the random forest prediction.
Overall, the random forest predictions better resemble the verification sample.

\begin{figure}
  \includegraphics[width=\hsize]{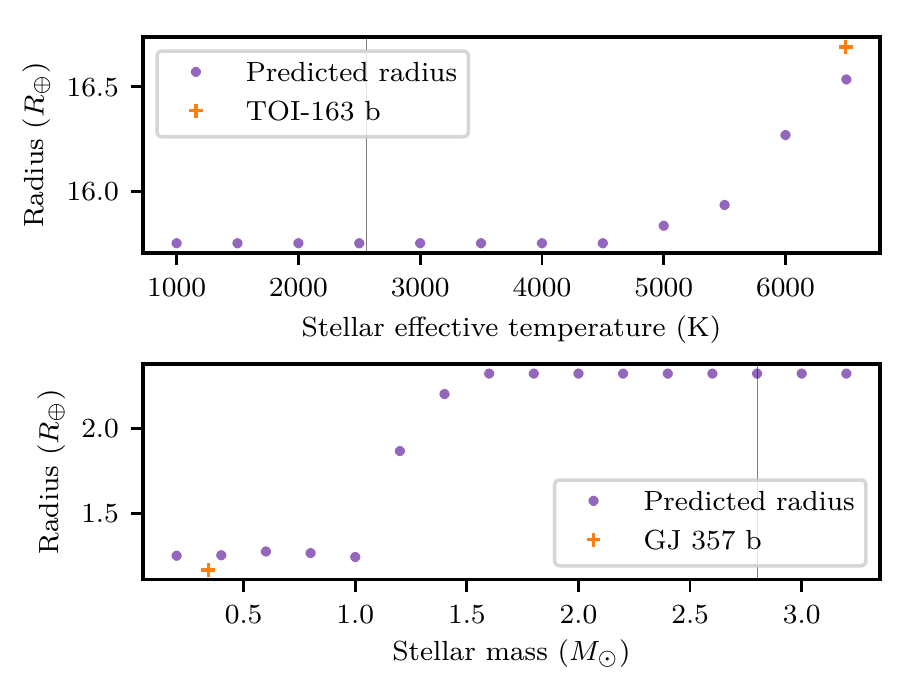}
  \caption{Top panel: Predicted radii as function of stellar effective temperature.
      Bottom panel: Predicted radii as function of stellar mass.
      The predicted radii are marked with purple dots, and the true radius and stellar effective temperature (or stellar mass) are marked with orange crosses.
      The grey line represents the extrema of the training parameter space.}
  \label{fig:extrapolate}
\end{figure}

\subsection{Limitation of the random forest model}
\label{limitation}

\begin{table*}
  \caption{Extrema values of planetary and stellar parameters in training set.}
  \label{table:extrema}
  \centering
  \begin{tabular}{c c c c c c c c c c}
                         & $M_{p}$         & $T_{eq}$   & a          & $R_*$     & $Teff_*$      & $M_*$      & $R_p$    \\
                         & ($M_\oplus$)   & (K)      & (AU)      & ($R_s$)     & (K)        & ($M_s$)      & ($R_\oplus$)  \\
  \noalign{\smallskip}
  \hline
  \noalign{\smallskip}
    Minimum value        & 0.0553       & 51       & 0.0111       & 0.117    & 2560.0     & 0.08      & 0.383 \\
    Maximum value        & 19750        & 2762     & 100          & 6.30     & 11361.0    & 2.80      & 23.37 \\
  \end{tabular}
\end{table*}

The random forest model is a data-driven technique that has the potential
to discover new correlations between parameters, but one of its limitations is the parameter space covered
by the training sample. Contrary to a linear relation for example,
where the extrapolation can predict values outside the range used to build the linear relation,
the random forest model is limited to the values present in the training sample.
Some parts of the parameter space covered by the exoplanets in the radial velocity sample
are not included in the parameter space of the training sample.
Table~\ref{table:extrema} details the minimum and maximum values of the parameters in the training sample.
For example, four exoplanets in the radial velocity sample have a planetary mass
that exceeds the maximum planetary mass in the training sample (\num{2e4}$M_\oplus$).
This implies that the random forest model is expected to extrapolate outside the training space.

To explain the behaviour of the model, we used two exoplanets
added to the database very recently: TOI-163 b \citep{kossakowski_2019}
and GJ 357 b \citep{luque_2019}, which are not part of our training sample.
We varied only one input parameter, such as stellar mass
while keeping the other parameters constant, and we predicted the planetary radius with the random forest model.
Figure~\ref{fig:extrapolate} presents the radius predictions
as the stellar effective temperature decreases below the minimum of the training sample (2560\,K)
and as the stellar mass exceeds the maximum of the training sample (2.8$\,M_\odot$).
The predicted radius of TOI-163 b stays constant for any stellar-effective temperature below 4500\,K,
even outside the parameter range. The predicted radius of GJ 357 b
also stays constant for any stellar masses above 1.6$\,M_\odot$, hence above 2.8$\,M_\odot$.

The random forest model extrapolates outside of the parameter space
by returning the radius' upper (or lower) bounds found when the training sample is used.
Of course, this is an important point to take into account when predicting the radius of an exoplanet with this model.
Outside the training parameter space, the estimated radii will not be reliable, since
no correlation can be predicted by the model.
The growing number of exoplanets with mass and radius measurements
(as well as the other parameters used in this model) implies that in the future the random forest model
could be trained again with a larger training sample, likely improving its predictive power.


\section{Conclusions}
\label{conclusion}

We built a random forest model which is able to predict the radius of exoplanets based on their mass,
their equilibrium temperature, and several stellar parameters.
The model covers a range of masses between \num{5.53e-2}$M_\oplus$ (Mercury) and \num{2e4}$M_\oplus$ (KOI-415\,b).
We find that the mass and the equilibrium temperature are the most important parameters in deriving the radius.
The gradient in equilibrium temperature, seen for the high mass planets, is well-estimated by the random forest model.
We compared our predicted radii with those measured and find a root-mean-squared error of 1.8$\,R_\oplus$.
Our model has a low bias, but a high variance that could be improved
as more exoplanets with mass and radius measurements are published.
One possible future step  towards developing this model is to include more stellar parameters,
such as stellar metallicity and stellar abundances,
even though the stellar abundances would restrict the number of exoplanets in the data set.

Random forests are a powerful algorithm for classification \citep{carliles_2010,ishida_2019}
and regression tasks. They might also be useful in the future
to predict stellar masses from other stellar parameters,
or to model other empirical relations
such as  the mass-metallicity-luminosity relation.

\subsection*{Acknowledgements}
This work was supported by Funda\c{c}\~ao para a Ci\^encia e a Tecnologia FCT/MCTES
through national funds (PIDDAC) and by FEDER -
Fundo Europeu de Desenvolvimento Regional funds through the COMPETE 2020 -
Programa Operacional Competitividade e Internacionalização (POCI)
by these grants:
UID/FIS/04434/2019;
UID/FIS/04434/2013\,\&\,POCI-01-0145-FEDER-007672;
PTDC/FIS-AST/32113/2017\&POCI-01-0145-FEDER-032113;
PTDC/FIS-AST/28953/2017\&POCI-01-0145-FEDER-028953.
J.B. acknowledges support by Funda\c{c}\~ao para a Ci\^encia e a Tecnologia (FCT)
through Investigador FCT contract of reference IF/01654/2014/CP1215/CT0003.
J.P.F. is supported in the form of work contract
funded by national funds through FCT with the reference DL 57/2016/CP1364/CT0005.

\bibliographystyle{aa}
\bibliography{MachineLearning}

%
%
\onecolumn
\begin{appendix}
  \section{Diagnostic plots}
  \label{diagnostic}

  \subsection{Learning and validation curves}
  \label{appendix:trans}

\begin{figure*}[h!]
      \includegraphics{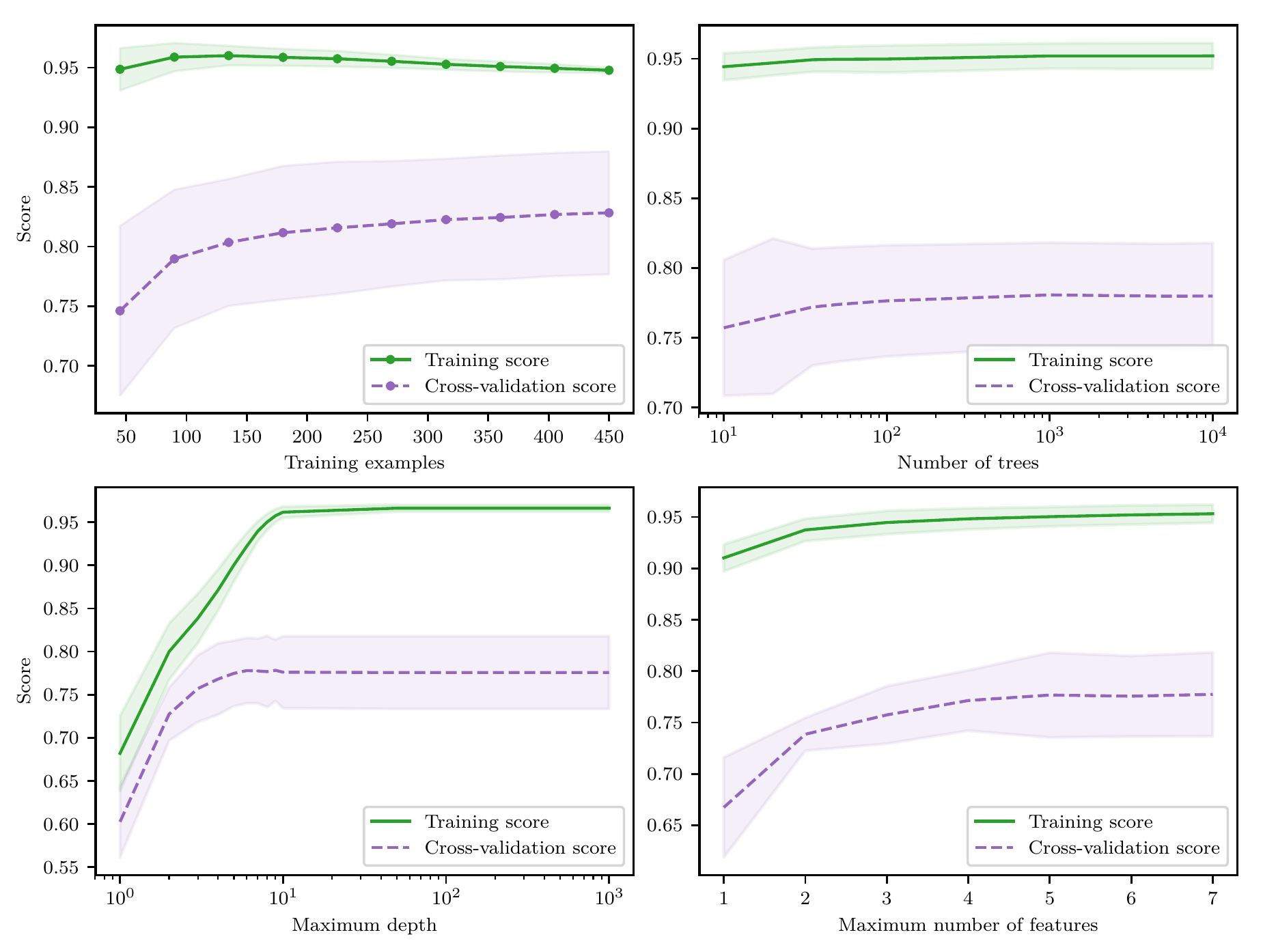}
      \caption{Learning and validation curves for random forest regressor.
        The scoring method is the $R^2$ score,
        the training set is represented in green, and the cross validation set in purple.}
      \label{fig:curves}
    \end{figure*}

   The learning and validation curves are a diagnostic tool of the random forest model.
   The first panel of Figure~\ref{fig:curves} shows the $R^2$ score of the training set is higher than for the cross validation set.
   The high $R^2$ score of 0.94 and
   the small error of 1.1$\, R_\oplus$ on the training set indicate
   the random forest model is able to describe the relation
   between the features and give an accurate prediction of the radius.
   This demonstrates that the model has a low bias.
   The high $R^2$ score of the training set
   and the lower score (around 0.82) of the validation set show that the model
   overfits the training set and does not generalise very well on the validation set.
   The gap between the two scores remains even when the whole training sample is used,
   which demonstrates that the curves do not converge.
   This behaviour indicates that the random forest model has a high variance,
   and it can be improved by constraining the hyperparameters:   for example, reducing the number of trees and their depth, or using feature bagging.
   Since we already implemented these methods to improve the output of the algorithm,
   another solution that could improve a model with high variance
   and low bias is to increase the training sample size.
   We would need more objects with mass and radius measurements
   so that the algorithm has more instances to capture the complexity of the relation.

\newpage
    
  \subsection{LIME explanation diagram}
  \label{appendix:lime}
\begin{figure*}[h!]
    \centering
    \includegraphics[width=0.5\textheight]{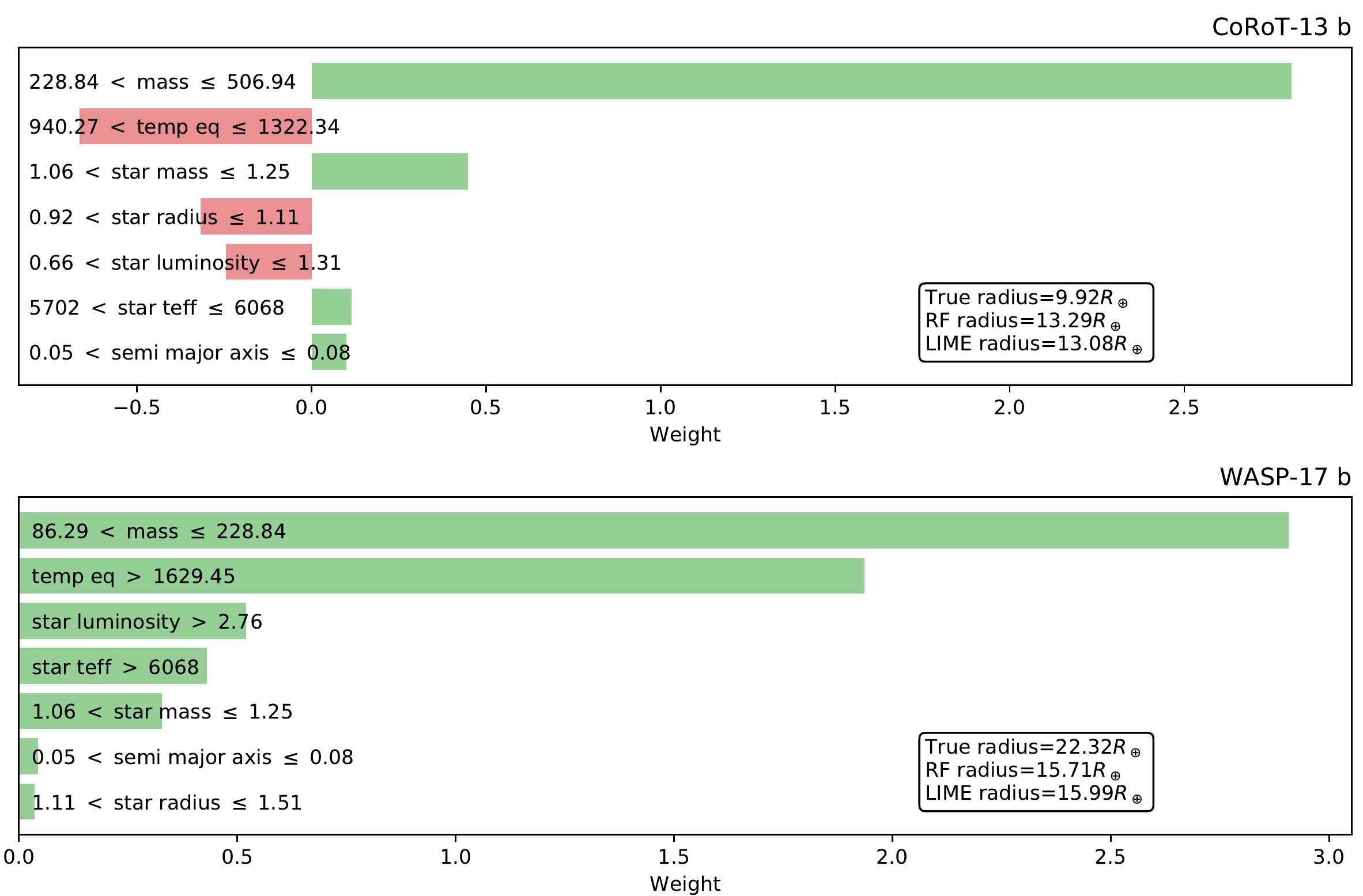}
    \includegraphics[width=0.5\textheight]{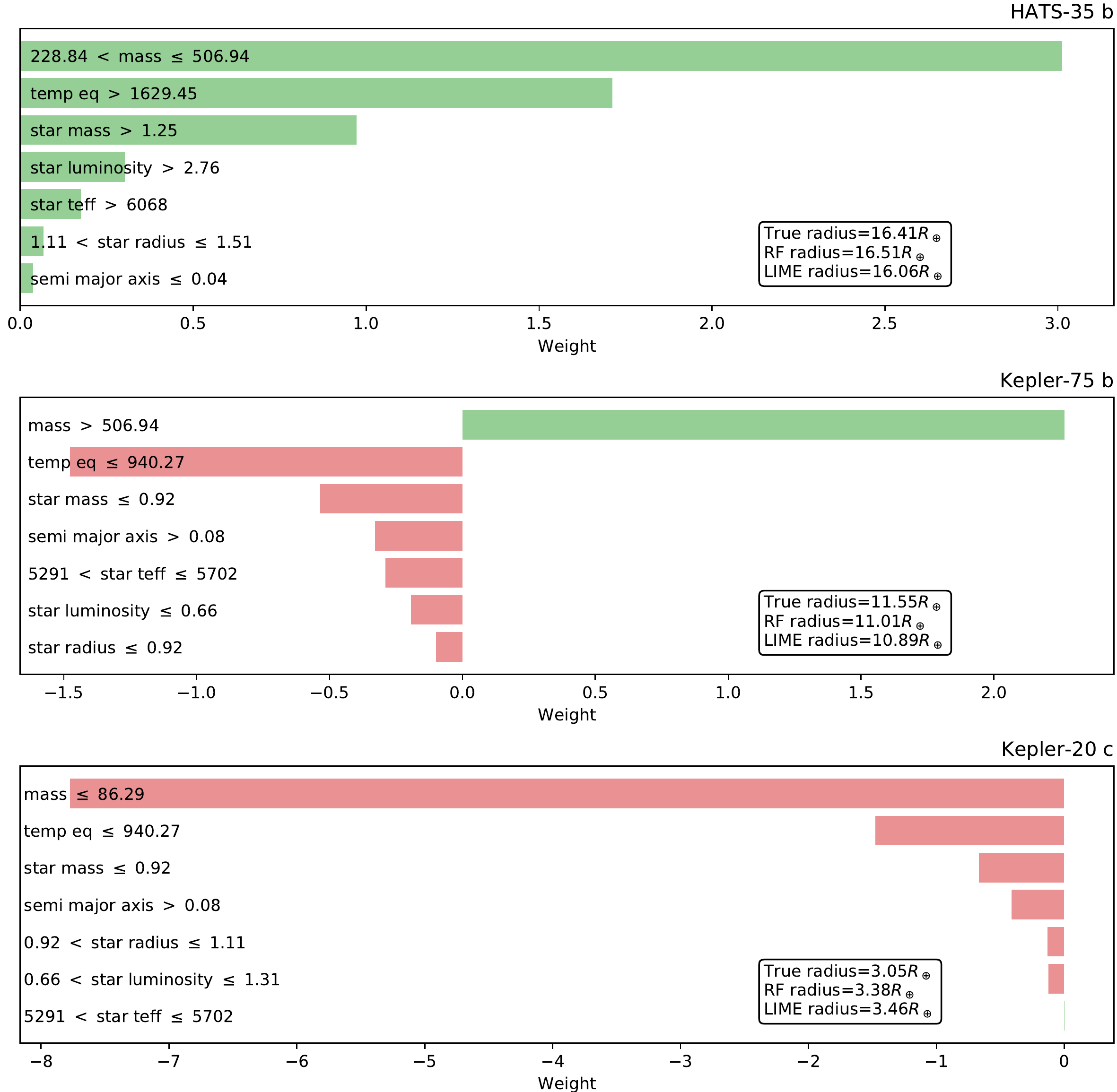}
    \caption{LIME explanations of the radius predictions.
      A positive weight is shown in green and a negative one in red for all input features.
       The predicted radii by the random forest model (RF radius) and by the LIME approximation (LIME radius)
       are compared to the true radius.}
      \label{fig:lime}
    \end{figure*}

  This appendix presents five exoplanets from the test set.
  HATS-35\,b, Kepler-75\,b, and Kepler-20\,c are well-predicted by the random forest model.
  But CoRoT-13\,b and WASP-17\,b are wrongly predicted.
  Figure~\ref{fig:lime} shows the local interpretation computed with LIME for each exoplanet.
  
\end{appendix}

\vfill
\eject
\end{document}